\documentclass[12pt]{article}
\usepackage{amsmath,amssymb,amsfonts,amsthm}

\title{On the ladder operators and nonclassicality of generalized coherent state associated with a particle in an infinite square well}

\author{{H R Jalali$^1$} and M K Tavassoly$^{1,2,3,*}$ \\
 \footnotesize{$^1$ Atomic and Molecular Group, Faculty of Physics, Yazd University, Yazd, Iran} \\
 \footnotesize{$^2$ Photonics Research Group, Engineering Research Center, Yazd University, Yazd, Iran} \\
 \footnotesize{$^3$ The Laboratory of Quantum Information Processing, Yazd University, Yazd, Iran}\\
 \footnotesize{$^*$ E-mail: mktavassoly@yazd.ac.ir}}

\begin{document}

\maketitle


\begin{abstract}
      In this paper the factorization method is used in order to obtain  the eigenvalues and eigenfunctions of a quantum particle confined in a one-dimensional infinite well. The output results from the mentioned approach allows us to explore an appropriate new pair of raising and lowering operators corresponding to the  physical system under consideration.  From the symmetrical considerations, the connection between the obtained ladder operators with $su(1,1)$ Lie algebra is explicitly established. Next, after the construction of Barut-Girardello and Gilmore-Perelomov representations of coherent states associated with the considered system,  some of their important  properties like the resolution of the identity including a few nonclassical features are illustrated in detail. Finally, a theoretical
      scheme for generation of the Gilmore-Perelomov type of coherent state via a generalized Janes-Cummings model is proposed.
\end{abstract}

{\bf Keywords:}

  Infinite potential well;  Factorization method;  Ladder operators;  Coherent state;  Nonclassical state.

 {\bf  PACS:} 42.50.Dv,42.50.-p


 \section{Introduction}\label{sec-intro}
  The time-independent form of Schr\"{o}dinger equation is written as:
  \begin{equation}\label{Hamiltoni0}
   H\psi=E\psi
 \end{equation}
 where $\psi$ and $E$ are the eigenfunction and  eigenenergy and $H$ represents  the Hamiltonian of the considered system defined as
  \begin{equation}\label{Hamiltoni1}
   H=\frac{p^{2}}{2m}+V(x),\qquad p=\frac{\hbar}{i}\frac{d}{dx}.
 \end{equation}
  As the first pedagogical examples which may be found in any text book of quantum mechanics, which has also enough applications in various fields of physics,  one may refer to a particle in an infinite well (in parallel with harmonic oscillator potential).
  But, while the creation and annihilation (ladder) operators associated with harmonic oscillator potential is a well-known matter in quantum mechanics, for the infinite potential well this is not explicitly known as is known for harmonic oscillator. This is our first purpose of the present paper.
  For a particle confined in one-dimensional box, the potential is defined as
 \begin{equation}\label{potential}
 V(x)=\left\{
        \begin{array}{ll}
          0 & \qquad  0\leq x \leq L \\
          \infty & \qquad x<0, \;x>L.
        \end{array}
      \right.
 \end{equation}
 The second-order differential equation in (\ref{Hamiltoni0}) for above $V(x)$ has been traditionally solved by using the power series method, and after applying the boundary conditions, i.e., $\psi(0)=\psi(L)=0$, one can obtain the exact form of the  eigenenergies and eigenfunctions. In this paper, we intend to solve the same equation by the factorization method through  which, with similar spirit of obtaining the ladder operators for the harmonic oscillator, we will be able to find out the generalized creation and annihilation operators for the  system under consideration. It is worth to mention that in contrast to the simple harmonic oscillator, the energy levels of a particle in one-dimensional box is not equally spaced.
 This observation  will make the problem more complicated. Anyway, we will try to find the appropriate pair of
 annihilation-creation operators corresponding to the mentioned
 system, by which we will be able to construct two distinct classes of
 coherent states.
In this respect, as is well-known, the increasing interest in
the coherent states \cite{cs1,cs2,cs3} is subject to the fact
that they are the main source of various nonclassical properties.
Nonclassical states have found their appropriate applications in various fields of physics in recent decades (for a useful and review text in coherent states see \cite{sissue} and references therein).
So, after exploring the appropriate
creation and annihilation operators we are lead to the construction of the corresponding coherent states by two well-known manners: algebraic and group theoretic method. At last, after demonstrating some of important properties of the introduced states, we investigate the nonclassicality features of them for completing our presentation. At last, by using a generalized Janes-Cummings model, a theoretical proposal for the generation of the Gilmore-Perelomov type of coherent state is presented.

We end this Introduction by recalling the fact that this potential, in a sense, has received a lot of attention in recent years.
Indeed, as an equivalent approach which is used for the infinite square well potential, we may recall the well-known SUSY quantum mechanics technique
(for instance see Refs. in \cite{newrefs}).
Also, the algebraic treatment for the system in hand,  can be found in \cite{refnew2}, and at last in the context of the construction of coherent states corresponding to  such system one may referee to Refs. \cite{refnew3}.
Altogether, it may clearly be seen that these works are all essentially different from the one has been presented here, either from the way to approach the problem
or from the final obtained results and presented discussions. Thus, to the best of our knowledge,
the approach to the treatment of the problem, the corresponding ladder operators, the associated coherent state of the type which is obtained in this paper,
in addition to the discussion on the resolution of the identity and the related nonclassicality signs may not be found
in the literature.

 In the continuation, at first, we briefly review the factorization method. In section 3, the method is used to solve the Schr\"{o}dinger equation for a particle in one-dimensional infinite square potential well. Based on the obtained results, we construct the creation and annihilation operators in section 4, by which we study the dynamical group corresponding to this system. In section 5,  the Barut-Girardello and Gilmore-Perelomov coherent states associated with a particle in the potential well
 are constructed. Then,  we demonstrate  (in section 6) some of the nonclassical properties of the introduced states like squeezing (first and second-order) and sub-Poissonian statistics, in addition to establishing the resolution of the identity corresponding to both classes of introduced states.
 A theoretical scheme for generation of the Gilmore-Perelomov type of coherent state is presented in section 7. Finally, section 8 deals with a summary and concluding remarks.

 \section{The factorization method: a brief review}\label{sec-nl}
  The factorization method was firstly introduced by Schr\"{o}dinger \cite{schrodinger1,schrodinger2,schrodinger3} and Dirac \cite{dirac} to avoid the use of cumbersome and lengthy mathematical calculations. It has been successfully applied to solve essentially the problems whose their exact solutions are exist. The formalism further developed by Infeld and Hull \cite{hull} and Green \cite{green}. The interest in the factorization method comes from the fact that we may discover the hidden symmetry of the quantum system through the establishment of an appropriate Lie algebra, which can be realized by the constructed ladder operators.

   The factorization method which is used in this paper  is equivalent to the SUSY treatment \cite{rodrigous}.
   Briefly, in SUSY one concerns with two Hamiltonians $H_-$ and $H_+$ which are  known as   SUSY partner Hamiltonians.
   The SUSY analysis allows one to generate a hierarchy of
   Hamiltonians \cite{sukumar}.
  In this method, the second-order differential operator $H$ in (\ref{Hamiltoni1}) is rewritten as the product of two
  first-order differential operators $a_{1}^{\dagger}$ and  $a_{1}$ plus a constant $E_{1}$, i.e.,
  \begin{equation}\label{susym}
  H_1=a_1^\dag a_1 +E_1.
   \end{equation}
  In  other words, rewriting the SUSY Hamiltonians  $H_-$ and $H_+$ respectively as  $H_1$ and $H_2$, and
  changing the subscripts in a suitable manner, we have (\ref{susym}) with its SUSY partner as
  $$H_2=a_1 a_1^\dag +E_1.$$
  Next, by factorizing $H_2$ one has
  $$H_2=a_2^\dag a_2 +E_2$$
  with its SUSY partner as
  $$H_3=a_2 a_2^\dag  + E_2.$$
  Continuing the above procedure leads to the
  generation of a hierarchy of Hamiltonians given by
 \begin{equation}\label{susym2}
  H_{j+1}=a_{j+1}^\dag a_{j+1} +E_{j+1} = a_{j} a_{j}^\dag  + E_{j},
 \end{equation}
  where $j=1, 2, 3, \cdots$.
 In general, these equations can have many solutions, but $a_j$ should be chosen in such a way that one gets the maximum value for $E_j$ (we will explain this issue in the continuation). In that case, the solution of equations (\ref{susym}) to (\ref{susym2}) will be unique. Also, recall that, $j$ is a counter indicator and so there is no indispensable reason for beginning it from $1$. Indeed, it can begin from 0, too, and we will be able to continue without any problem. However, due to the outlined example which we will  work with it in the next sections of the paper (a particle in an infinite square well) it seems to be reasonable to start  with 1  (notice that  $\psi_{j}(y)\sim \sin (j y)$).

 Now, upon using the following theorem we can find eigenfunctions and eigenvalues of the system Hamiltonian \cite{ohanian}.\\
 \textbf{Theorem}: Assume that equation (\ref{susym}) and  equations (\ref{susym2}) with real constants $E_{j}$ are satisfied. Also, suppose that there exists a null eigenfunction $|\xi_{j}\rangle$ with zero eigenvalue for each $a_{j}$, i.e.,
 \begin{equation}\label{Help}
   a_{j}|\xi_{j}\rangle=0.
 \end{equation}
 Then,\\
 a) $E_{j}$ is the $j$th eigenvalue of the Hamiltonian.

 b) Corresponding eigenfunction of the Hamiltonian $H$ is of the form
 \begin{equation}\label{eigenfunction}
   |\psi_{j}\rangle=c_{j}a_{1}^{\dagger}a_{2}^{\dagger}...a_{j-1}^{\dagger}|\xi_{j}\rangle
 \end{equation}
 where $c_{j}$ is a normalization constant may be determined as:
 \begin{equation}\label{normalization}
   |c_{j}|=[(E_{j}-E_{j-1})(E_{j}-E_{j-2})...(E_{j}-E_{1})]^{-1/2}.
 \end{equation}
There are three important points in the theorem:

i) This method applies only to the discrete energy spectra.

ii) In the selection of $a_j$, one should choose it in such a way that $E_{j}$ gets the maximum value  \cite{ohanian}.
Indeed, it is an obvious fact which comes from the assumption $a_j|\xi_{j}\rangle=0$. For more detail, if one supposes that $E_j$ and $a_j$ are changed, whereas $E_{j-1}$ and $a_{j-1}$ remain fixed, then:
 $$\langle\xi_j|\delta E_j|\xi_j\rangle = -\langle\xi_j|\delta (a_j^\dagger a_j)|\xi_j\rangle =
 \langle\xi_j|(-\delta a_{j}^{\dagger} a_j-a_{j}^{\dagger} \delta a_j )|\xi_j\rangle,$$
 $$ = -\langle \xi_j|\delta a_{j}^{\dagger} (a_j|\xi_j \rangle)-(\langle \xi_j| a_{j}^{\dagger})  \delta a_j |\xi_j\rangle = 0. $$

iii) The theorem doesn't give any information about the form of $a_{j}$. Moreover, evidently the form of $a_{j}$ depends explicitly
 on the particular potential $V(x)$ in entered in (\ref{Hamiltoni1}). Anyway, whereas $a_{1}^{\dagger}a_{1}$ is equal to $H$, except for a constant, $a_{1}$ must contain a linear momentum term to be consistent with the kinetic energy part of the Hamiltonian. Accordingly, $a_{j}$ can be expressed as
 \begin{equation}\label{annihil-opera}
   a_{j}=\frac{1}{\sqrt{2m}}(p+if_{j}(x))
 \end{equation}
 where $p$ is the momentum operator and $f_{j}(x)$ is a real and differentiable unknown function of $x$ by which equation (\ref{susym}) is satisfied. These operators are not hermitian $(a_{j}^{\dagger}=\frac{1}{\sqrt{2m}}(p-if_{j}(x))\neq a_{j})$, however using the commutation relation $[f_{j}(x),p]=i\hbar \frac{df_{j}(x)}{dx}$, their  hermitian product $a_{j}^{\dagger}a_{j}$ can be written as follows
 \begin{eqnarray}\label{a dag a}
   a_{j}^{\dagger}a_{j}=\frac{p^{2}}{2m}+g_j(x)
    \end{eqnarray}
where the real function $g_j(x)$ is  defined as
$g_j(x) \equiv \frac{f_{j}^{2}(x)}{2m}+\frac{\hbar}{2m}\frac{df_{j}(x)}{dx}$.

\section{A particle in one-dimensional infinite potential well}\label{sec-n2}
 Now, we are going to search for the eigenenergies and eigenfunctions of our system, i.e. a particle in an infinite well potential by the factorization method. At first, with considering equation (\ref{a dag a}) for $j=1$ and using (\ref{susym}) we arrive at the following equation for $f_{1}(x)$
\begin{equation}\label{Riccati-eq}
   \frac{df_{1}(x)}{dx}=-\frac{f_{1}^{2}(x)}{\hbar}-\frac{2mE_{1}}{\hbar}
 \end{equation}
  which is a simple Riccati differential equation. The above equation can be easily solved using the change of variable $f_{1}(x)=\sqrt{2mE_{1}}\cot(y)$, from which we finally have:
 \begin{equation}\label{f-func}
   f_{1}(x)=\sqrt{2mE_{1}}\cot\left(\frac{\sqrt{2mE_{1}}}{\hbar}(x-a)\right)
 \end{equation}
 where $a$ is an integration constant. Possible selections for $E_{1}$ and $a$ may be restricted by cotangent function and since $0 \!\leq \! x \! \leq \! L$, the cotangent function must be finite in this allowed range. Of course we should get the maximum value of $E_{1}$ and it will correctly be obtained if one of the cotangent singularities be in $x=0$ and the other in $x=L$. Hence, we choose the point $x=0$ as the singularity point of $f_{1}(x)$,
 \begin{equation}\label{a=0}
 f_{1}(0)=\infty \;\;\;\;\;\;\rightarrow \;\;\;\;\;\; a=0
 \end{equation}
 with substituting $a=0$ in (\ref{f-func}) and choosing $x=L$ as the other singularity of $f_{1}(x)$
\begin{equation}\label{limit}
 f_{1}(L)=\infty \;\;\;\;\;\;\rightarrow \;\;\;\;\;\; \frac{\sqrt{2mE_{1}}}{\hbar}L=\pi.
\end{equation}
 As a result, with substituting (\ref{a=0}) and (\ref{limit}) in (\ref{f-func}) and using equation (\ref{annihil-opera}),  $a_{1}$ can be determined as follows
 \begin{equation}\label{a for particle}
   a_{1}=\frac{1}{\sqrt{2m}}\left(p+i\frac{\pi\hbar}{L}\cot(\frac{\pi}{L}x)\right)
 \end{equation}
 and from (\ref{limit}) the ground state energy will be achieved as:
\begin{equation}\label{E 1}
E_{1}=\frac{\pi^{2}\hbar^{2}}{2mL^{2}}.
 \end{equation}
 Now, after finding the form of the function $f_{1}(x)$ as the ground state, in order to obtain all of the eigenstates, we consider
 \begin{equation}\label{fj}
 f_{j}(x)=c_{j}\cot(b_{j}x)
 \end{equation}
 and make use of recurrence relation (\ref{susym2}) with the restriction $0\leq b_{j} \leq \frac{\pi}{L}$, then one has
 \begin{eqnarray}\label{recu-part}
   \frac{1}{2m}[p^{2}&-&c_{j+1}b_{j+1}\hbar+c_{j+1}(c_{j+1}-b_{j+1}\hbar)\cot^{2}(b_{j+1}x)]+E_{j+1} \nonumber\\
 &=& \frac{1}{2m}[p^{2}+c_{j}b_{j}\hbar + c_{j}(c_{j}+b_{j}\hbar)\cot^{2}(b_{j}x)]+E_{j}
 \end{eqnarray}
The equation (\ref{recu-part}) will be true, if
\begin{equation}\label{b}
   b_{j+1}=b_{j}\;\;\;\;\rightarrow\;\;\;\;b_{j}=b_{1}=\frac{\pi}{L}
 \end{equation}
 \begin{equation}\label{c}
   c_{j+1}(c_{j+1}- b_{j+1}\hbar)=c_{j}(c_{j}+ b_{j}\hbar)\;\;\;\;\rightarrow\;\;\;\;c_{j+1}=\left\{
                                                                                               \begin{array}{ll}
                                                                                                 -c_{j} \\
                                                                                                 c_{j}+\frac{\pi\hbar}{L}
                                                                                               \end{array}
                                                                                             \right.
 \end{equation}
 \begin{equation}\label{E}
   2mE_{j+1}-c_{j+1}b_{j+1}\hbar=2mE_{j}+c_{j}b_{j}\hbar
   \;\;\;\;\rightarrow\;\;\;\;E_{j}=\frac{c_{j}^{2}}{2m}.
 \end{equation}
 From the solutions of (\ref{c}) the second answer namely
 $c_{j+1}=c_{j}+\frac{\pi\hbar}{L}$ provides the maximum value of
 $E_{j+1}$, as we require, from which one obtains
 \begin{equation}\label{cj}
   c_{j}=\frac{j\pi\hbar}{L}.
 \end{equation}
Finally, from (\ref{E}) and (\ref{cj}) we can derive the eigenvalues as
\begin{equation}\label{E kamel}
E_{j}=\frac{j^{2}\pi^{2}\hbar^{2}}{2mL^{2}},\;\;\;\; j=1,2,3,... \;.
\end{equation}
With substituting $c_{j}$ and $b_{j}$, respectively from (\ref{cj}) and (\ref{b}) in (\ref{fj}) and using the obtained result in (\ref{annihil-opera}), $a_{j}(x)$ will be explicitly obtained
\begin{equation}\label{ajx}
  a_{j}=\frac{1}{\sqrt{2m}}\left( p + i\frac{j\pi\hbar}{L}\cot(\frac{\pi}{L}x)\right).
 \end{equation}
 The equation (\ref{E kamel}) gives all the eigenvalues if there was one
 answer for arbitrary $j$ in (\ref{Help}). Thus, we require to solve
 \begin{equation}\label{diff Help-func}
   a_{j} \xi_{j}(x)=0
 \end{equation}
 from which one obtains
 \begin{equation}\label{calcu Help-func}
   \frac{d\xi_{j}(x)}{dx}=j\frac{\pi}{L}\cot(\frac{\pi}{L}x)\xi_j(x).
 \end{equation}
 Solving the above equation for $\xi_{j}(x)$ one easily reads
 \begin{equation}\label{Help-func}
 \xi_{j}(x)=\left(\sin(\frac{\pi}{L}x)\right)^{j}
  \end{equation}
 and finally by using (\ref{eigenfunction}) the normalized eigenfunctions of a particle in box will be obtained  as
 \begin{equation}\label{func-particle}
   \psi_{j}(x)=\sqrt{\frac{2}{L}}\sin(\frac{j\pi}{L}x).
 \end{equation}
Therefore, the evaluated results in (\ref{E kamel}) and (\ref{func-particle}) are our final solutions of the considered eigenvalue problem.

 \section{Ladder operators associated with infinite potential well}\label{sec-n3}

Notice that $a_j$ operators in (24) and their corresponding conjugates do not play the role of
lowering and raising operators for the energy levels of a particle in a
box, i.e., they are not our required ladder operators. But, as we will see in the continuation of the paper, we
will still able to construct a pair of annihilation-creation operators after a
change of variable as  $y=\frac{\pi}{L}x$ in the above results.
Indeed, after some tries, we acquire the following operators, yielding our above requirement
\begin{equation}\label{annihilition-y}
A_{j}\equiv\frac{d}{dy}+j\cot(y)
 \end{equation}
 and $A_{j}^{\dagger}$
 \begin{equation}\label{Ajd}
A_{j}^{\dagger}\equiv -\frac{d}{dy}+j\cot(y).
 \end{equation}
 By acting the operator $A_{j}^{\dagger}$ on $\psi_{j}(y)$ we have
\begin{eqnarray}\label{recu-particle}
   A_{j}^{\dagger}(\psi_j( y))&=&\sqrt{\frac{2}{L}}[j\cos(j y)+j\cot(y)\sin(j y)] \nonumber\\
   &=&\frac{j}{\sin(y)}\sqrt{\frac{2}{L}}\sin[(j+1)y]=\frac{j}{\sin(y)}\psi_{j+1}(y).
 \end{eqnarray}
 Consequently
 \begin{equation}\label{sin a}
   \sin(y)A_{j}^{\dagger}\psi_{j}(y)=j\psi_{j+1}(y).
 \end{equation}
 Similarity, by applying $A_{j}$ on  $\psi_{j}(y)$ we obtain
 \begin{equation}\label{sin aa}
   \sin(y)A_{j}\psi_{j}(y)=j\psi_{j-1}(y).
 \end{equation}
 By the way, keeping in mind the above relations, at this stage it is possible to introduce the ladder operators which can
 really raise and lower any eigenfunction (or corresponding energy level) of our physical system
 \begin{equation}\label{Lpm}
   K_\pm=\pm\sin(y)\frac{d}{dy}+\cos(y)K_0
 \end{equation}
 where we have defined the (Hermitian) number operator $K_0$
 that satisfies the eigenvalue equation
 \begin{equation}\label{number opera}
   K_0|j\rangle = j|j\rangle, \;\;\;\;
   \psi_{j}(y)=\langle y|j \rangle.
 \end{equation}
 Briefly, it is easy to check that the generalized ladder operators $K_\pm$ in (\ref{Lpm}) possess the following required properties
 \begin{equation}\label{proper Lpm}
   K_\pm|j\rangle = j|j\pm1\rangle, \;\;j=1,2,3,..., \qquad K^\dag_{\pm} =K_\mp\; .
 \end{equation}
 A word seems to be necessary about the ladder operators for arbitrary solvable quantum system (and corresponding generalized coherent states).
 In fact, the number operator dependent of the ladder operators are usual in this typical treatments such as SUSYM quantum mechanics etc (for instance see \cite{daood}).
 The operators $K_\pm$ and $K_0$ justify the following
 commutation relations
 \begin{equation}\label{Lie}
   [K_{-},K_{+}]=2K_0, \;\;\;\;
   [K_0,K_\pm]=\pm K_\pm.
 \end{equation}
 As is well known, according to (\ref{Lie}), in the infinite dimensional Hilbert space spanned by
 $|j\rangle, j=1,2,3,\cdots $, the set of operators $\{K_-,K_+,K_0\}$ are the generators of $su(1,1)$ Lie algebra.
 In addition, the system Hamiltonian can be simply expressed as
 \begin{eqnarray}\label{Hamiltony Lpm}
   H&=&\frac{\pi^{2}\hbar^{2}}{2mL^{2}}(K_{+}K_{-}+K_0)=\frac{\pi^{2}\hbar^{2}}{2mL^{2}}(K_{-}K_{+}-K_0)\\ \nonumber
   &=& \frac{\pi^2\hbar^2}{4 m L^2} \{K_+, K_-\}
 \end{eqnarray}
 where $\{.\; , .\}$ stands for the anticommutation relation.
 All of the expressed forms for the Hamiltonian in (\ref{Hamiltony Lpm}) satisfy the eigenvalue equation $H|j\rangle = E_{j}|j\rangle$ according to (\ref{E kamel}).
 The operator equation (\ref{Hamiltony Lpm}) may also be considered as an explicit definition for $K_0$-operator. It is worth to notice that the last relation in (\ref{Hamiltony Lpm}) seems to be similar to that of the harmonic oscillator, i.e., $H_{HO}=\hbar \omega \{b, b^\dagger\}$, where $b, b^\dagger$ are the standard bosonic annihilation and creation operators which together with number operator $b^\dag b$ construct the Weyl-Heisenberg Lie algebra.

   \section{The construction of generalized coherent states}\label{sec-n4}

 Due to the recently increasing interest in the coherent states and nonclassical states in quantum optics, we are now in the position to produce the coherent states for our considered system. Generally there are three main approaches for the construction of coherent states, i.e., annihilation operator eigenstate (algebraic consideration), displacement operator (group theoretical consideration) and dynamical consideration. Only for harmonic oscillator all of the three manners lead to a unique state known as standard or canonical coherent state $|z\rangle=\exp(-|z|^2/2) \sum_{n=0}^\infty \frac{z^n}{\sqrt n} |n\rangle$. After finding the physical realization of coherent state through the discovery of the direct relation of laser and coherent state, all of the mentioned approaches are vastly generalized in literature. Henceforth, now which we are being equipped with the ladder operators of the infinite well potential we are able to produce the generalized coherent states associated with the system by generalization of the annihilation operator eigenstate and displacement operator manners.
 \begin{itemize}
 \item {\it Annihilation operator eigenstate:}
 The Barut-Girardello coherent state (algebraic method) is defined as the right eigenstate of the annihilation operator, i.e.,
\begin{equation}\label{Lm alpha}
   K_{-}|\alpha \rangle_{BG}=\alpha |\alpha \rangle_{BG}.
 \end{equation}
 To deduce the explicit form of  $|\alpha \rangle_{BG}$, with the traditional
 method we can write
 \begin{equation}\label{alpha is sigma}
   |\alpha \rangle_{BG}=\sum_{j=1}^{\infty} c_{j}|j\rangle.
 \end{equation}
 Then, we set it in (\ref{Lm alpha}) and upon using (\ref{proper Lpm}) finally arrives us at Barut-Girardello type of coherent states as follows
 \begin{equation}\label{Barut}
   |\alpha \rangle_{BG}=N(|\alpha|^{2})\sum_{j=0}^{\infty}
   \frac{\alpha^{j}}{(j+1)!}|j+1\rangle
 \end{equation}
 where the factor $N(|\alpha^{2}|)$ can be calculated by the
 normalization condition,  i.e., $\langle \alpha |\alpha \rangle =1$ results in
 \begin{equation}\label{normal Barut}
   N(|\alpha|^{2})=\left(\sum_{j=0}^{\infty}
   \frac{|\alpha|^{2j}}{[(j+1)!]^2}\right) ^{-\frac 1 2}.
  \end{equation}
 \item {\it Displacement operator coherent state:}
 On the other hand, keeping in mind the $su(1,1)$ symmetry algebra of the generators as in (\ref{Lie}), we are able to construct the Gilmore-Perelomov coherent states associated with the particle in an infinite  potential well as follows \cite{Miri_Tavassoly_Scripta} (group theoretical method)
  \begin{eqnarray}\label{displace oper}
   |\alpha \rangle_{GP}&=&D(\xi)|1\rangle=\exp(\xi K_{+}-\xi^{\ast} K_{-})|1\rangle \nonumber \\
   &=&\exp(\alpha K_{+})(1-|\alpha|^{2})^{K_0} \exp(-\alpha^{\ast} K_{-})|1\rangle
 \end{eqnarray}
 with $\alpha=\frac{\xi}{|\xi|}\tanh(\xi)$. Notice that the disentanglement theorem has been used in obtaining the final result in above equation. Straightforward calculations lead one to the normalized form of Gilmore-Perelomov coherent states, i.e.,
 \begin{equation}\label{Gilmore}
   |\alpha \rangle_{GP}=\sqrt{1-|\alpha|^{2}}\sum_{j=0}^{\infty}\alpha^{j}|j+1\rangle.
 \end{equation}
 Now after finding out the two distinct classes of coherent states, we are ready to investigate some of nonclassical properties which are of most importance in quantum optics studies. But, before paying attention to this subject, we should establish the resolution of the identity for the above introduced states.

\item  {\it Resolution of the identity}: Any generalized  coherent state such as $|\alpha\rangle_{BG}$ should satisfy the resolution of the identity which is defined as follows
 \begin{equation}\label{Resolution}
   \int d^{2}\alpha w(|\alpha^{2}|)|\alpha\rangle_{BG} {}_{BG}\langle \alpha|=\hat{I}=\sum_{j=0}^{\infty}|j+1\rangle \langle j+1|
 \end{equation}
 where $w(|\alpha^{2}|)$ is a suitable weight function should be found. Substituting (\ref{Barut}) in (\ref{Resolution}) and using the change of variables $\alpha=re^{i\theta}$ and $r^{2}=x$ lead us to the following relation
 \begin{equation}\label{Resolution partic}
   \pi\sum_{j=0}^{\infty}|j+1\rangle \langle j+1|\int_{0}^{\infty}w(x)N^{2}(x)x^{j}dx=[\Gamma(j+2)]^{2}
 \end{equation}
 where $\Gamma(j+2)$ is the gamma function. This is indeed an inverse moment problem which can be solved by the well-known methods such as Mellin transform \cite{klauder}. It can  be easily checked that the required weight function which can satisfy the above integral equation is of the form
 \begin{equation}\label{weight func}
   w(x)=\frac{2xK_{0}(2\sqrt{x})}{\pi N^{2}(x)}
 \end{equation}
 where $K_{0}(2\sqrt{x})$ is the Bessel function of second kind.\\
 The Gilmore-Perelomov states according to (\ref{Gilmore}), are formally as harmonious coherent states. The
 resolution of the identity for such states has been discussed in \cite{manko,sudur}.

 Therefore, adding the obvious nonorthogonality of the two obtained states in (\ref{normal Barut}) and (\ref{Gilmore}) to the presented discussion in this section, we conclude the over-completeness relation for both classes of introduced coherent states.

 \end{itemize}

 \section{Nonclassical features of the introduced states}\label{sec-n5}
 \begin{itemize}
 \item  \textit{Normal squeezing}: In order to examine the quantum fluctuations of the quadratures of the field \cite{walls}, following the well-known generalization of the field quadratures $x, p$ \cite{roy33, Chithiika}, we introduce the hermitian field operators
 $X_{1}=\frac{K_{-}+K_{+}}{2}$ and $Y_{1}=\frac{K_{-}-K_{+}}{2i}$. The squeezing parameters can be defined as follows
 \begin{equation}\label{squeeze}
   s_{\gamma}=\frac{(\Delta\gamma)^{2}}{\sqrt{\frac{1}{4}|\langle [X_{1},Y_{1}]\rangle|^{2}}}-1,\;\; \gamma=X_{1},Y_{1}
 \end{equation}
 where $(\Delta \gamma)^2= \langle \gamma^2\rangle -  \langle \gamma\rangle^2$.
 A state is squeezed in $X_{1}$ or $Y_{1}$ if it satisfies the inequalities $-1\!<\!s_{X_{1}}\!<\!0$ or $-1\!<\!s_{Y_{1}}\!<\!0$, respectively.
 \\
 Figure 1 shows that the squeezing for Barut-Girardello coherent states occurs only in $Y_{1}$ component in all real space. Similarity, from figure 2 we understood that this nonclassical feature for Gilmore-Perelomov are similarly hold in $Y_{1}$ quadrature. All figures are symmetrical around $\alpha=0$ and the minimum and maximum of these parameters occur in approximately $\alpha=0$.
 \\
 \item \textit{Amplitude-squared squeezing}:  This parameter is defined in terms of Hermitian operators
 $X_{2}=\frac{K_{-}^{2}+K_{+}^{2}}{2}$ and $Y_{2}=\frac{K_{-}^{2}-K_{+}^{2}}{2i}$.
 The latter generalizations originate from the proposal of Hillery \cite{hillery} which was firstly introduced in connection with the square of the complex amplitudes of the electromagnetic field. The squeezing conditions in $X_{2}$ or $Y_{2}$ are respectively given by $-1\!<\!S_{X_{2}}\!<\!0$ or $-1\!<\!S_{Y_{2}}\!<\!0$, where $S_{X_{2}}$ and $S_{Y_{2}}$ defined as follows
 \begin{equation}\label{amplitude}
 S_{\ell}=\frac{(\Delta\ell)^{2}}{\sqrt{\frac{1}{4}|\langle [X_{2},Y_{2}]\rangle|^{2}}}-1,\;\; \ell=X_{2},Y_{2}.
 \end{equation}
 In figures 3 and 4 amplitude-squared squeezing  has been plotted against $\alpha$, respectively for Barut-Girardello and Gilmore-Perelomov coherent states.
  According to the displayed figures, it is clear that this parameter for both of introduced coherent states occurs in $Y_{2}$ direction, too.\\
  \item  \textit{Quantum statistical properties}: For establishing the statistical properties and specially to examine the sub-Poissonian statistics as another nonclassicality sign \cite{Davodovich}, we calculate the Mandel parameter \cite{mandel}, which can be generalized as
 \begin{equation}\label{mandel}
   Q=\frac{\langle K_{+}^{2}K_{-}^{2} \rangle - \langle K_{+}K_{-} \rangle^{2}}{\langle K_{+}K_{-} \rangle}-1.
 \end{equation}
 The state for which $Q\!=\!0$, $Q\!<\!0$ and $Q\!>\!0$ respectively corresponds to the Poissonian (standard coherent states) sub-Poissonian (nonclassical states) and super-Poissonian (classical states) statistics.
 \\
 In figures 5 and 6 Mandel parameter has been plotted versus $\alpha$, respectively for Barut-Girardello and Gilmore-Perelomov coherent states. By comparing the two figures it is deduced that, while Barut-Girardello coherent states show this nonclassical feature in all considered regions, the Gilmore-Perelomov coherent states possess this property in a finite region of the allowed space.
 \end{itemize}

\section{Generation of Gilmore-Perelomov type of coherent state}
Recently one of us with his coauthor have proposed  a theoretical generation scheme for producing any class of Gilmore-Perelomov type of $SU(1, 1)$ coherent states  \cite{Miri_Tavassoly_Scripta} (and their superpositions).
It is readily found that the proposal can help us for the generation of our GP coherent state, too.
To achieve this purpose, we consider the atom-field interaction
in which a two-level atom interacts with a quantized single-mode
cavity field via an intensity-dependent coupling
together with an  external classical field.
In the resonance condition and under the rotating-wave approximation,
the dynamical evolution of the mentioned system may be appropriately described by the Hamiltonian in the nonlinear Janes-Cummings regime
as follows ($\hbar=1$):
\begin{equation}\label{H}
   H =\lambda(\sigma_{-} e^{i\varphi} + \sigma_{+} e^{-i\varphi})+ \Lambda( A^\dagger \sigma_{-} + A \sigma_{+}),
\end{equation}
 where $\sigma_{-} = |g\rangle\langle e|$ and $\sigma_{+} = |e\rangle\langle g|$ are the atomic lowering and raising  operators. Also, the parameters $\lambda$ and $\Lambda$ (assuming that $\Lambda \gg  \lambda$) are
respectively the coupling coefficients of the atom with classical and quantized cavity fields, $A$ and $A^\dagger$ are the $f$-deformed
ladder operators and $\varphi$ is the phase of classical field. In  \cite{Miri_Tavassoly_Scripta} it is assumed that the operators $A$, $A^\dag$ and their commutators satisfy the $su(1, 1)$ Lie algebra (this does not occur for any arbitrary $f$-deformed ladder operators).
In the present paper we deal really with the generators of such a specific group, therefore we begin with the following Hamiltonian instead
\begin{equation}\label{H}
   H =\lambda(\sigma_{-} e^{i\varphi} + \sigma_{+} e^{-i\varphi})+ \Lambda( K_+ \sigma_{-} + K_- \sigma_{+}),
\end{equation}
where $K_+$ and $K_-$ (with $K_0$) constitute the generators of the $su(1, 1)$ algebra.
It is demonstrated that, in the strong classical field regime
$(\Lambda\gg\lambda)$, the time evolution operator of the system
has been defined by \cite{zou}:
\begin{equation}\label{U-t}
   U(t)= R^\dagger T^\dagger(t) U_{eff} T(0) R,
\end{equation}
where the operators $R$, $T$ and $U_{eff}$ are given by:
\begin{equation}\label{R}
   R=\exp{\left[\frac{\pi}{4} (\sigma_{+}-\sigma_{-})\right]} \exp{\left(\frac{i\varphi}{2}\sigma_{z}\right)},
\end{equation}
\begin{equation}\label{T}
   T(t)=\exp{\left(i\lambda \sigma_{z} t\right)},
\end{equation}
\begin{equation}\label{Ueff}
   U_{eff}=\exp{\left[-\frac{i\Lambda t}{2} (K_+ e^{-i\varphi} + K_- e^{i\varphi})\sigma_{z}\right]} .
\end{equation}
Now, demanding the generation of our Gilmore-Perelomov states corresponding to a particle in an infinite well, the initial atom-field state
is denoted by
\begin{equation}\label{state0}
  |\psi(0)\rangle = \left(\frac{|e\rangle +|g\rangle}{\sqrt{2}}\right) |1\rangle
\end{equation}
which means that the cavity is initially prepared in the ground  state of the field and the atom in a
superposition of excited and ground states with equal weights.

It is a straightforward matter to obtain the final state of the system, following the mentioned procedure of \cite{Miri_Tavassoly_Scripta}, which yields the final state of the atom-field system at time $t$ given by:
\begin{eqnarray}\label{state4}
 |\Psi(t)\rangle=\frac{1}{\sqrt{2}} \left( e^{-\frac{i\varphi}{2}} e^{-i\lambda t} \cos\frac{\varphi}{2}  | \alpha \rangle_{GP} + i e^{-\frac{i\varphi}{2}} e^{i\lambda t} \sin\frac{\varphi}{2}  |- \alpha \rangle_{GP} \right) |e\rangle\nonumber \\
  +\frac{1}{\sqrt{2}} \left(e^{\frac{i\varphi}{2}} e^{-i\lambda t} \cos\frac{\varphi}{2}  | \alpha \rangle_{GP} - i e^{\frac{i\varphi}{2}} e^{i\lambda t} \sin\frac{\varphi}{2}  |- \alpha \rangle_{GP} \right) |g\rangle,
 \end{eqnarray}
  where we have set $\alpha \doteq - \frac{i\Lambda t}{2} e^{-i\varphi}$.
  The final result  $|\Psi(t)\rangle$ is a general superposition of Gilmore-Perelomov  coherent states of $SU(1,1)$ group.
 The state of quantized cavity field may be determined after doing a measurement on the initial state of  the atom.
 For instance, if the atom is detected in excited or ground state, the state of the field will be collapsed respectively to:
\begin{equation}\label{state5}
 |\Psi^{+}\rangle =\frac{N^{+}}{\sqrt{2}} \left(e^{-\frac{i\varphi}{2}} \cos\frac{\varphi}{2}  | \alpha \rangle_{GP} + i e^{-\frac{i\varphi}{2}} \sin\frac{\varphi}{2} | -\alpha \rangle_{GP}\right),
\end{equation}
or
\begin{equation}\label{state6}
 |\Psi^{-}\rangle =\frac{N^{-}}{\sqrt{2}} \left(e^{\frac{i\varphi}{2}} \cos\frac{\varphi}{2}  | \alpha \rangle_{GP} - i e^{\frac{i\varphi}{2}} \sin\frac{\varphi}{2} | -\alpha \rangle_{GP}\right),
\end{equation}
where $N^{\pm}=\sqrt{2}$  and $\lambda t = 2k\pi$.   equations (\ref{state5}) and (\ref{state6}) show that the cavity field has been arrived at a
combination of $SU(1,1)$ coherent state.

Now notice that, the expansion coefficients in (\ref{state5}) and
(\ref{state6}) are $\varphi$-dependent.  For instance, in particular, with
selecting $\varphi = 2 \pi$ ($\varphi =\pi$), regardless of the
atomic detection, the state of the field collapses to
$ | \alpha \rangle_{GP}$ ($ | -\alpha \rangle_{GP}$),
i.e.,
$SU(1,1)$ coherent states may be generated. \\
Although out of the scope of the present paper it is worth mentioning the  possible generation of  even or odd superposition of
Gilmore-Perelomov  $SU(1,1)$ coherent states by mean of the proposal in \cite{Miri_Tavassoly_Scripta}, too.

    \section{Summary and conclusion}\label{sec-n5}
In this article we obtained the eigenvalues and eigenfunctions of a single particle in a one-dimensional infinitely deep square  potential well by the factorization method. Then, appropriate generalized creation and annihilation operators are obtained based on the used method and derived results. Then, we have explored a symmetrical realization of the dynamical group; it is demonstrated  that these operators can constitute a $SU(1,1)$ group. Next, we have constructed the explicit form of two distinct classes of coherent states, i.e., Barut-Girardello and Gilmore-Perelomov coherent states for the  system under our consideration. The resolution of the identity and so the over-completeness relation for both states are illustrated in detail. Finally, in view of the importance of nonclassical states in recent decades in both theoretical and experimental aspects, we were succeeded in highlighting some of the nonclassical properties such as first and second-order squeezing and sup-Poissonian statistics for our introduced states, numerically.
This is while the standard coherent states of harmonic oscillator possess neither of the nonclassicality features outlined  in this paper.
At last, via the nonlinear Janes-Cummings model, a theoretical scheme for generation of the Gilmore-Perelomov type of the outlined coherent state  is proposed.
Also, one can go further to produce even and odd superposition of both states in addition to excited coherent states which may be done elsewhere.

We end the concluding remarks with mentioning the fact that there exists a general way for the construction of generalized coherent state associated with any solvable quantum system with  nondegenerate  discrete spectrum (such as the  one considered in this article) known as Gazeau-Klauder type of coherent states \cite{GK, GGK}.  But, while our approach is rather new and novelty, the results (the obtained ladder operators, the corresponding generalized coherent states and their properties) are all essentially different and distinguishable.
Moreover, the Gazeau-Klauder coherent state corresponding to a particle in an infinite well is very similar to P\"{o}shl-Teller potential has been adequately studied in \cite{GK_Posh} and so should not be considered here.


{\bf Acknowledgement:}  The authors are grateful to thank the referees for their useful suggestions which improved the contents of the paper. Also,
 thanks to  PhD students: S R Miry and M J Faghihi in Atomic and Molecular Physics in Yazd University for their helps in preparing the final form of the paper.

\newpage
{\bf Figure Captions:}

  {\bf FIG. 1}:  Plot of squeezing parameters, $s_{X_{1}}$ (up diagram) and $s_{Y_{1}}$ (down diagram) against $\alpha$ for Barut-Girardello coherent states.

  {\bf FIG. 2}:  Plot of squeezing parameters, $s_{X_{1}}$ (up diagram) and $s_{Y_{1}}$ (down diagram) against $\alpha$ for Gilmore-Perelomov coherent states.

  {\bf FIG. 3}:  Plot of amplitude-squared squeezing parameters, $S_{X_{2}}$ (up diagram) and $S_{Y_{2}}$ (down diagram) against $\alpha$ for Barut-Girardello coherent states.

  {\bf FIG. 4}: Plot of amplitude-squared squeezing parameters, $S_{X_{2}}$ (up diagram) and $S_{Y_{2}}$ (down diagram) against $\alpha$ for Gilmore-Perelomov coherent states.

  {\bf FIG. 5}: Plot of Mandel parameter for Barut-Girardello coherent states  versus $\alpha$.

  {\bf FIG. 6}: Plot of Mandel parameter for Gilmore-Perelomov coherent states  versus $\alpha$.


\newpage

\begin{thebibliography}{999}


\bibitem{cs1}   S Twareque Ali,  J-P Antoine,   J-P Gazeau, Coherent States, Wavelets and Their Generalization,
Springer, New York, 2000.

\bibitem{cs2} J R  Klauder,  B-S Skagerstam, Coherent States, Applications in Physics and
Mathematical Physics, Word Scientific, Singapore, 1985.


\bibitem{cs3}  J-P  Gazeau,  Coherent States  in Quantum Physics, Weinheim: Wiley-VCH 2009.

\bibitem{sissue} See the papers appeared in: 2012 {\it J. Phys. A: Math. and Theor.}
  45  No. 24  Special issue on coherent states: mathematical and physical aspects.

\bibitem{newrefs}  M M Nieto, L M Simmons 1978 {\it Jr., Phys. Rev. Lett.} {\bf  41}  207;
 D J Fernandez, {\it AIP Conf. Proc.} {\bf 809} (2006) 80;
 R F Fox, M H Choi  2001 {\it Phys. Rev. A} {\bf 64} 042104;
 L Dello Sbarba,  V Hussin 2007 {\it J. Math. Phys.} {\bf 48}  012110;
 S  Kuru, J Negro 2008 {\it Annals of Phys.} {\bf 323} 413;
 F Cooper, A Khare, U. Sukhatme 1995  {\it Phys. Rep.} {\bf 251} 267;
 A Khare 2005  {\it AIP Conf. Proc.}  {\bf 744} 133;

\bibitem{refnew2}  M M  Nieto, L M  Simmons 1979 {\it Jr., Phys. Rev. D} {\bf 20}  1332;
C Quesne 1999 {\it J. Phys. A: Math. Theor.} {\bf  32} 6705;

\bibitem{refnew3} M M Nieto, L M  Simmons 1979 {\it Jr., Phys. Rev. Lett.} {\bf 41} 207;
 A H  El Kinani, M  Daoud 2002 {\it Int. J. Mod. Phys. B} {\bf 16} 3915;
 D J Fernandez 2006 {\it AIP Conf. Proc.} {\bf 809} 80;
 D J  Fernandez, V Hussin, O Rosas-Ortiz 2007 {\it J. Phys. A: Math. Theor.} {\bf 40}  6491;

\bibitem{schrodinger1} E Schr\"{o}dinger 1941 {\it Proc. Roy. Irish Acad. Sect.} A {\bf 46}  9.

\bibitem{schrodinger2} E Schr\"{o}dinger 1941  {\it Proc. Roy. Irish Acad. Sect. A} {\bf  46}  183.



\bibitem{schrodinger3} E Schr\"{o}dinger 1941   {\it Proc. Roy. Irish Acad. Sect. A} {\bf 47}  53.

\bibitem{dirac}  P M  Dirac, The principles of quantum mechanics,  Oxford University Press, Oxford 1958.

\bibitem{hull} I Infeld, T E Hull 1951   {\it Rev. Mod. Phys.}  {\bf 23}  21.

\bibitem{green} H S  Green,  Matrix method in quantum mechanics,  Barnes and Noble, New York 1968.

\bibitem{rodrigous} R de Lima Rodr´ýguez  2002  "{\it The quantum mechanics SUSY algebra: an introductory review}", Preprint hepth/0205017.

\bibitem{sukumar} C Sukumar 1985 {\it J. Phys. A: Math. Gen.} {\bf 18} L57;  C Sukumar 1985 {\it J. Phys. A: Math. Gen.} {\bf 18} 2917.


\bibitem{ohanian}  H C Ohanian, Principles of quantum mechanics,  Prentice Hall. Englewood Cliffs, NJ.  1990.

\bibitem{daood} M Daoud and V Husssin 2002  {\it J. Phys. A: Math. Theor.} {\bf 35} 7381;
                 L Dello Sbarba and V Hussin 2007 {\it J. Math. Phys.}  {\bf 48} 012110;
                 S Cruz y Cruz, S Kuru and J Negro 2008 {\it Phys. Lett. A} {\bf 372} 1391.

\bibitem{Miri_Tavassoly_Scripta}  S R  Miry, M.K. Tavassoly  2012   {\it Phys. Scr.}  {\bf 85} 035404.

\bibitem{klauder}  J R Klauder, K A Penson, J-M. Sixdeniers 2001  {\it Phys. Rev. A}  {\bf 64}  013817.

\bibitem{manko}   V I Man'ko, G Marmo, E C G Sudarshan, F Zaccaria 1997 {\it Phys. Scr.}   {\bf 55} 528.

\bibitem{sudur} E C G  Sudarshan  1993 {\it Int. J. Theor. Phys.} {\bf 32} 1069.

\bibitem{walls} D F Walls  1983  {\it Nature}  {\bf 306}  141.

\bibitem{roy33} B Roy, P Roy 2002 {\it Phys. Lett.} A {\bf 296}  187.

\bibitem{Chithiika} V Chithiika Ruby, S Karthiga and M Senthilvelan 2013  {\it J. Phys. A: Math. Theor.} {\bf 46} 025305.

\bibitem{hillery} M Hillery 1987 {\it Opt. Commun.} {\bf 62}  135.

\bibitem{Davodovich} L Davodovich 1996 {\it Rev. Mod. Phys.} {\bf 68} 127.

\bibitem{mandel} L Mandel 1979  {\it Opt. Lett.}  {\bf 4}  205.

\bibitem{zou} X Zou, K Pahlke and W Mathis 2004 {\it Phys. Rev. } A {\bf 69} 015802.

\bibitem{GK} J-P Gazeau, J R Klauder 1999  {\it J. Phys. A} {\bf 32} 123.

\bibitem{GGK} F Yadollahi,  M K Tavassoly 2011   {\it Opt. Commun.} {\bf 284}  608.

\bibitem{GK_Posh}  J-P Antoine,  J-P Gazeau,  J R Klauder, P Monceau,  K A  Penson 2001 {\it J. Math. Phys.} {\bf  42} 2349.

\end{thebibliography}
\end{document}